# What Practitioners Really Think About Continuous Software Engineering: A Taxonomy of Challenges

Muhammad Zohaib
Computer Science Department, Govt. College University Faisalabad, Punjab-Pakistan
zohiab@gcuf.edu.pk

**Abstract**
The Continuous software engineering is a collaborative software development environment which offers the continues development and deployment of quality software project within short time. The Continuous software engineering practices are not yet mature enough, and the software organizations hesitate to adopt it. This study aims: (1) to explore the Continuous software engineering challenges by conducting systematic literature review (SLR) and to get the insight of industry experts via questionnaire survey study; (2) to prioritize the investigated challenges using fuzzy analytical hierarchy process (FAHP). The study findings provides the set of critical challenges faced by the software organizations while adopting Continuous software engineering and a prioritization based taxonomy of the Continuous software engineering challenges. The application of FAHP is novel in this research area as it assists in addressing the vagueness of practitioners concerning the influencing factors of Continuous software engineering. We believe that the finding of this study will serve as a body of knowledge for real world practitioners and researchers to revise and develop the new strategies for the successful implementation of Continuous software engineering practices in the software industry.

## 1. Introduction

The software organization is continuously looking the better ways to develop good quality software with a significant return on investment. There is a dramatic change in the software development approaches 'from traditional waterfall to agile paradigm' over the years. Currently, the software organizans rapidly changing the development environment in terms of shortening and continuous development and release cycles, using the state-of-the-art development approach, namely as Continuous software engineering (development and operations) (Erich, Amrit et al. 2017, Forsgren 2018, Leite, Rocha et al. 2019). The Continuous software engineering environment has been adopted and accepted over a few years, and still, it is a lack of widely accepted definition. In this study, we have coated a recent and more comprehensive definition of Continuous software engineering defined by (Leite, Rocha et al. 2019) "Continuous software engineering is a collaborative and multidisciplinary effort within an organization to automate continuous delivery of new software versions while guaranteeing their correctness and reliability."

In order to shotren the development life cyle and to employ the continuous delivery process, the software development community increasingly adopting the practices of Continuous software engineering (O'Hanlon 2006, Feitelson, Frachtenberg et al. 2013, Chen 2015). (Bai, Li et al. 2018) stated that the production of quality projects and the in-time delivery is the critical aspect of software business organizations, and the continuous software engineering is essential to achieve such goals, which assist in adoptingthe Continuous software engineering paradigm. (Balalaie, Heydarnoori et al. 2016), (Sharma and Coyne 2017) mention that the developers, operators, customers, and quality assurance teams continuously collaborated for delivery, reduce time, and attain market opportunities. They further indicated that the high-quality project production and delivery, rapid and timely entertainment of requirements changes, and reduced development time accelerated the acceptance of Continuous software engineering in the software industry. The IBM elaborated that the Continuous software engineering is a business-oriented software development and delivery methodology as it consists of the lines of business, practitioners, managers, and suppliers(Chen 2015, Sharma and Coyne 2017). Various established digital giants like WebEx, McAfee, CISCO, Netflix, and Amazon are already using the Continuous software engineering practices to deliver the perfect fitting customer-centric software solution in the international market(Zhu, Bass et al. 2016). The software industry, especially the mediocre software organizations, faced various complexities while adopting Continuous software engineering practices. Aiming to implement the Continuous software engineering practices in software development organizations successfully, the mainstream research body has motivated to assists the practitionerin developing the new techniques and tools(Chen, Kazman et al. 2015, Callanan and Spillane 2016). Due to the increasing demand for Continuous software engineering in the software industry, currently, it becomes the hot research topic.

Besides the significance and criticality of Continuous software engineering in the software industry, little empirical research has been carried out to fix the complications faced by the practitioners. Hence, the importance of Continuous software engineering in software industry motivated us to conduct a compressive study to explore and analyze the challenges that are critical for the successful execution of Continuous software engineering paradigm. To address the study objective, firstly, the literature review was performed to explore the challenges reported by the academic researcher and were further validated them with real world practitioners via questionnaire survey. Secondly, the fuzzy analytical hierarchy process (FAHP) approach was applied to prioritize the investigated challenges concerning to their criticality for Continuous software engineering paradigm. Various researchers already used this technique in other software engineering domain. For example, (Khan, Shameem et al. 2019)used the FAHP to rank the success factors of software process improvement. (Yaghoobi 2018) used the FAHP approach to rank the success factors of software project management. (Bozbura, Beskese et al. 2007) prioritize the measurement indicators of human capital using FAHP. (Shameem, Kumar et al. 2020) developed analyses the success factors of agile software development process. We believe that the in-depth review and analysis of the Continuous software engineering challenges will help the industry experts to revise their strategies and develop new roadmaps for the success and progression of Continuous software engineering execution in the software industry. The proposed research question of this study are:

RQ1: What challenges of Continuous software engineering paradigm are reported in the literature?
RQ2: Are the Continuous software engineering challenges reported in the literature related to real-world practices?
RQ3: How the investigated challenges be prioritized?
RQ4: What would be the taxonomy of the investigated challenges?

## 2. Background of Continuous software engineering

The evolution of the rapid revolution in information technology causes the transformation of development tools and techniques. The business firms are highly motivated to transform their working environment from manual to digital form as the automation increases productivity and maintains the consistency of product quality, which significantly increases the demand for software systems. To meet the market demand, the software organization continually looking at the active development approaches to develop and deliver the quality software's orders within time and budget(Sebastian, Ross et al. 2017). (Dörnenburg 2018) indicated that to meet the market demand and to address the technological transformation effectively, the software organization needs to adopt new and efficient software development approaches. By seeking this technological revolution and market trend, the traditional software development approaches (like Waterfall, Spiral, etc.) were replaced by the agile paradigm (i.e., Scrum and Kanban, etc.). The production and operational process are stressful, as manual processing is error-prone and causes a delay in feedback(Humble and Farley 2010). Therefore, to meet the current flows in the software industry, Continuous software engineering is the new and more efficient software development paradigm, which is based on the agile practices and operational aspects. The Continuous software engineering approach gave a complementary set of agile methods that assists to efficiently and continuously release the developed features in a shortened life cycle.

Initially, the term Continuous software engineering has ambiguity in its interoperation as part of the software community consider it a job opportunity that requires both skills, i.e., development and operation(Senapathi, Buchan et al. 2018). This ambiguity addressed by mainstream research by interpreting its actual meaning as Continuous software engineering is a development environment in which both development and operational teamwork with close collaboration (Roche 2013, Jabbari, bin Ali et al. 2016, Senapathi, Buchan et al. 2018). In Continuous software engineering, the distinct silos for developers and operators still exist; the operational team is responsible for the management of modification during production and in-service levels [48], other-side the development staff are accountable for the continuous development of new features to attain the required business goals. Both teams have their independent tools, process, and knowledge bases. This mechanism allows the development staff to push new features into production continuously, and the operational teams attempt to operate the latest version and highlight the modification to maintain the consistency in project quality and other non-functional requirements(Humble and Farley 2010). To address the flows between development and operation teams, an automated pipeline is needed to be considered (Woods 2016). (Humble and Farley 2010) stated that "the humble advocates for an automated deployment pipeline, in which any software version committed to the repository must be a production-candidate

version."(Humble 2010) underlined that the automation process defines a path that allows the development and auto testing, and the tested feature of the software is sent to the production by pressing the button. (Callanan and Spillane 2016) emphasized the continuous delivery and stated the deployment pipeline as a Continuous software engineering platform.

## 3. Research Design
In this paper, the research was design in three different steps:
**Step-1:** Identified the Continuous software engineering challenges reported in the literature using a systematic literature review.
**Step-2:** To get the insight of the industry practitioners concerning to the Continuous software engineering challenges, the questionnaire survey approach was used.
**Step-3:** Rank the identified list of challenges using the fuzzy AHP approach.

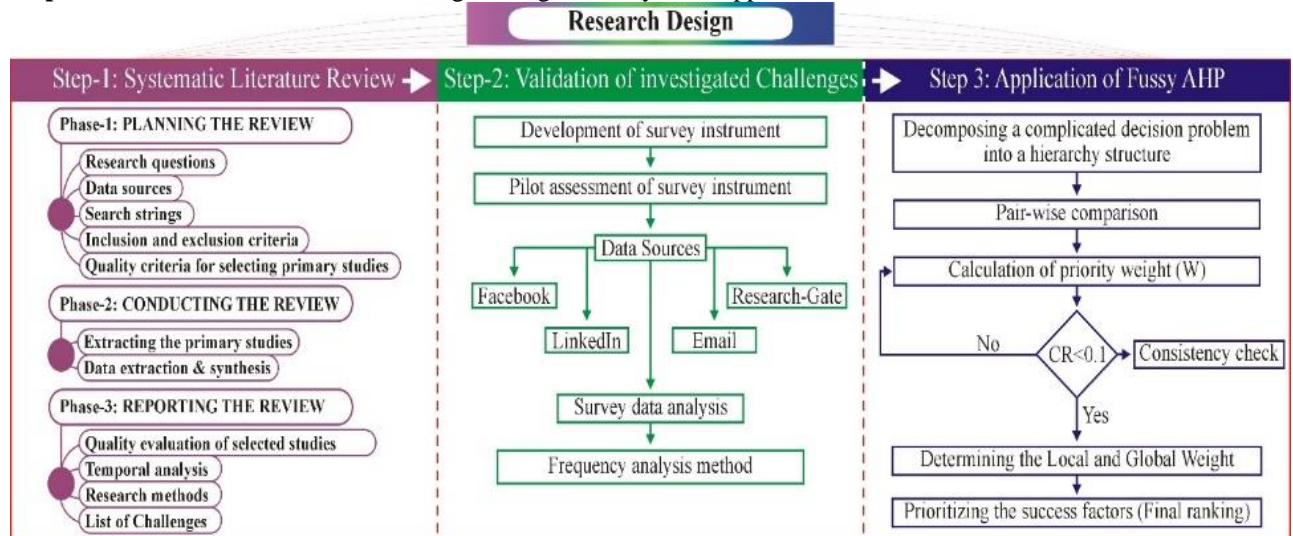

Figure 1: Adopted research design

### 3.1 Systematic Literature Review (SLR)
An SLR approach has been applied to collect and review the literature related to the study objectives. An SLR give the more comprehensive and valid results compare with informal literature review. The SLR guidelines proposed by (Kitchenham and Charters 2007) were considered to extract the potential literature related to the study objectives. According to (Kitchenham and Charters 2007), the SLR includes three core phases: "planning the review," "conducting the review," and "reporting the review." The developed SLR protocols are explained in the subsequent section and diagrammatically indicated in Figure 1.

### 3.1.1 Planning the review
To conduct the literature review, the following review protocols were developed:
Research questions:
The literature review was performed to identify the Continuous software engineering challenges reported in the literature. Though, the developed RQ is:
[RQ1] What challenges of Continuous software engineering paradigm are reported in the literature?

*Data collection source:*
For the collection of most relevant literature concerning to research objective, the selection of appropriate digital databases are important. Therefore, for the selection of digital repositories, we have fellow the suggestions of (Chen, Ali Babar et al. 2010) and (Zhang, Babar et al. 2011). The selected repositories include:

 I.     "IEEE Xplore (http://ieeexplore.ieee.org)"
 II.    "ACM Digital Library (http://dl.acm.org)"
 III.   "Springer Link (http://link.springer.com)"
 IV.    "Wiley Inter-Science (www.wiley.com)"

V. "Science Direct (http://www.sciencedirect.com)"
VI. "Google Scholar (http://scholar.google.com)"
VII. "IET Software (https://digital-library.theiet.org)"

*Search string:*
An appropriate search string play a key role to extract the potential literature from selected data sources. To extract the literature form the selected databases, we have develop a search string collecting the key terms and their substitutes by considering the guidelines of Qazi gold standards(White, Glanville et al. 2001) and (Zhang, Babar et al. 2011). To formulate the complete search string, we used the Boolean "OR" and "AND", as presented below:

("barriers" OR "obstacles" OR "hurdles" OR "difficulties" OR "impediments" OR "hindrance" OR "Concerns" OR "techniques" OR "tools," OR "methods," OR "process" OR "evaluation") AND ("Continuous software engineering" OR "Development and Operation," OR "Continues development and operation."

*Initial inclusion criteria:*
The protocols were developed to decide the inclusion of literature collected from the selected databases. The inclusion protocols were designed by following the existing studies (Niazi, Mahmood et al. 2016) and (Inayat, Salim et al. 2015). (1) The paper published in a journal, conference, or book chapter. (2) The article should explain the challenges of Continuous software engineering implementation. (3) Study results based on empirical data sets. (4) The paper should have a clear motivation for Continuous software engineering adoption. (5) Selected literature should be in English language.

*Initial exclusion criteria:*
We have further developed the protocols to exclude the literature collected from databases initially. The exclusion criteria were developed by following the guidelines of (Niazi, Mahmood et al. 2016), (Inayat, Salim et al. 2015) and (Akbar, Sang et al. 2019). (1) the most completed study from a similar research group was considered. (2) The paper should provide detail description of Continuous software engineering implementation. (3) The study that not related to the study objective. (4) The study is full or regular paper. (5) The literature review studies were not considered.

*Study quality assessment (QA):*
The QA assessment process was performed to determine that how the selected literature effective to answer the research objective. The QA process is carried out by using the guidelines of (Kitchenham and Charters 2007). For the QA process, the five-questions were developed (Table 1) and evaluated using the Likert scale given in Table 1. Similar criteria are adopted by various existing studies (Inayat, Salim et al. 2015, Niazi, Mahmood et al. 2016, Akbar, Sang et al. 2019). The detailed score of QA is given in Appendix-A.

Table 1: Checklist for QA

| QA No. | Checklist Questions | Likert scale |
|---|---|---|
| QA1 | Does the used research approach address the research questions? | "Yes=1, Partial=0.5, NO=0" |
| QA2 | Does the study discuss the challenges of Continuous software engineering? | "Yes=1, Partial=0.5, NO=0" |
| QA3 | Does the study have a clear motivation for Continuous software engineering implementation? | "Yes=1, Partial=0.5, NO=0" |
| QA4 | Is the collected data related to Continuous software engineering practices execution? | "Yes=1, Partial=0.5, NO=0" |
| QA5 | Are the identified results related to the justification of the research questions? | "Yes=1, Partial=0.5, NO=0" |

### 3.1.2 Conducting the review
*Final study selection:*
The three different ways were used to collect the literature. Firstly, 6 studies were collected manually by exploring Research-Gate. Secondly, the selected databases were explore by executing the search string and 688 studies were extracted. Therefore, for the final refinement of studies, the tollgate approach developed by (Afzal, Torkar et al. 2009) was adopted. By steps of tollgate approach (Figure 2), 54 studies were selected. Furthermore, we have performed the forward and backward snowballing on the reference list of selected studies, and 19 studies were selected. To conclude, 78 articles were considered for data extraction. All the studies were also assessed concerning the QA, and the results are given in Appendix-A. Each selected study is presented as "PS" to present its use as an SLR study.

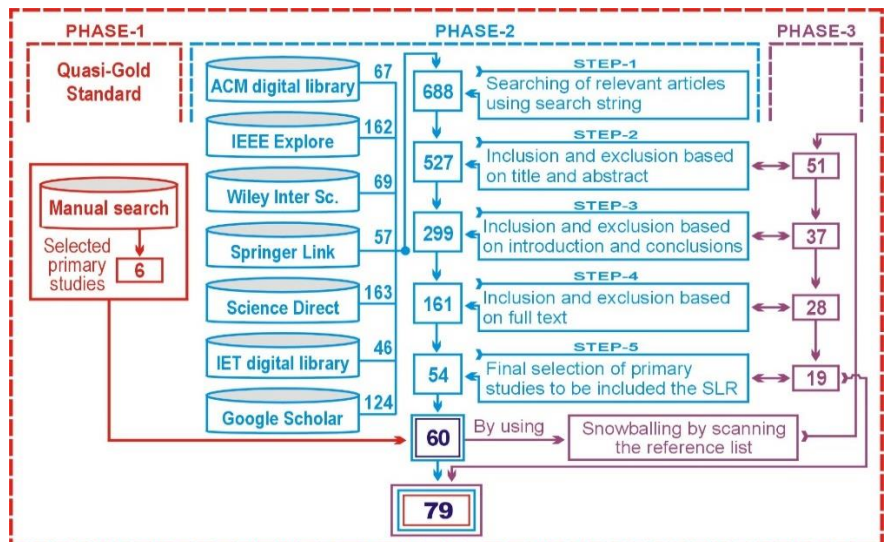

Figure 2: Refinement of formal studies

*Data extraction and synthesis:*
The selected studies (Figure 2) were carefully reviewed for data extraction with correspondence to the study research objective. The first two authors of this study were continuously involved in the data extraction process, and author numbers three and four validate the extracted data. Initially, the statements, main themes, concepts, and Continuous software engineering challenging factors were obtained from the selected studies. We then synthesized the collected data into compact statements and formed the final 20 challenging factors of Continuous software engineering implementation in the software industry.

There may be a biasness between the study findings. Though, the "inter-rater reliability test" (Afzal, Torkar et al. 2009) was performed. We have requested the four external experts for the participation in validation process. They randomly selected the 12 studies and performed the data extraction process. Based on the findings of study authors and external experts, we have calculated the "non-parametric Kendall's coefficient of concordance" (W)(Hallgren 2012). The value of W=1 renders the complete agreement, and W=0 indicates the complete disagreement. The results of W=0.84 p=0.003 shows an agreement between the investigation of study authors and external experts. This indicated that the study findings are unbiased. The used code is given in this link:*https://rdrr.io/cran/DescTools/man/KendallW.html*.

### 3.1.3 Reporting the review
*Quality of selected studies:*
The quality assessment of the selected studies shows how the selected literature is effective to answer the research question of this study. According to the accumulative results of the QA process shows that more than 70% of studies score ≥ 70%. The detail QA results are presented in Appendix-A. We have to use the 50% score as a threshold value.

*Publication years and used research approaches in selected studies:*
We extract the publication years of the selected studies to determine the frequency of publication of Continuous software engineering related literature. The analysis indicated that the chosen set of studies was published from 2013 to 2019, and this shows a growing trend in the frequency of publication in recent years. Hence, this renders that Continuous software engineering is an important and attractive research area of mainstream research body. In addition, we also extracted the adopted research methodologies in the selected studies. The results shows that the selected studies respectively adopted "questionnaire survey" (QS, 18%), "case study" (CS, 35%), "grounded theory"

(GT, 17%), "content analysis" (CA, 5%), "action research" (AR, 9%) and "mixed-method" (MM, 16). Therefore, we observed that CS is the most commonly used research approach.

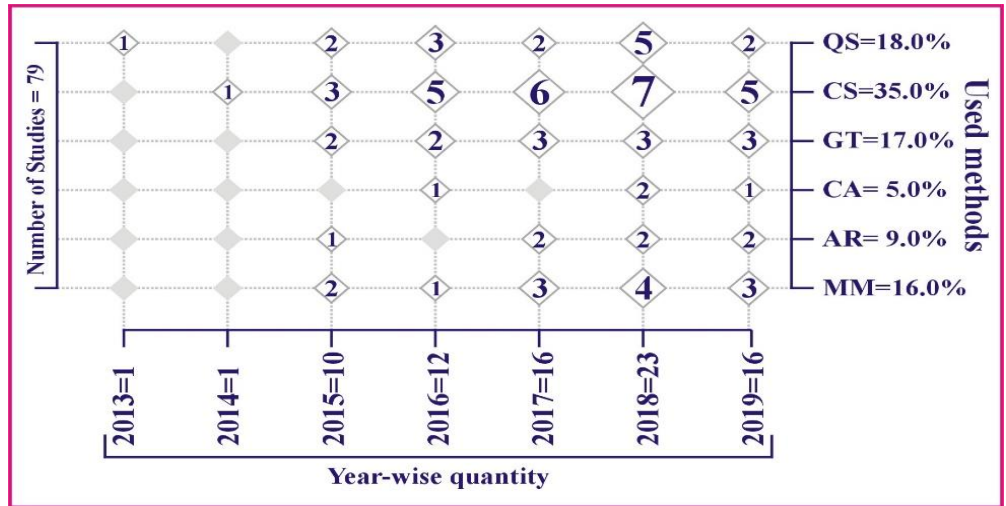

**Figure 3:** Publication years and adopted research approaches based methodologies

### 3.2 Empirical study
The questionnaire survey is a suitable way to collect the most potential data from the dispersed and targeted population.
According to (Kitchenham and Pfleeger 2003) the assortment data collection method is based on "available data collection resources," "controlling mechanism of selected approach," and "skill to operate the variable of interest." To collect the representative sample, the observation methods are hard(Akbar, Sang et al. 2019, Akbar, Sang et al. 2019). Hence, we have used the questionnaire method to answer the research question of this study, as it isan effective data collection way from dispersed population.

3.2.1 Survey instrument development
We developed a questionnaire survey to get the feedback of real-world industry practitioners. To develop the questionnaire survey, the Google Form (i.e.,"docs.google.com/forms") is used. The questionnaire survey was consists of three main sections, which include; (1) the first section contains the queries related to the respondents bibliographic information. (2) the second section includes the list of challenges extracted from literature (3) the third it contains the survey participants inputting the additional challenging factor which is not enlisted in the closed-ended section.

### 3.2.2 Pilot study of questionnaire survey
After the development of the questionnaire, we have conducted a pilot assessment with the academic and industry experts to determine the understandability of the developed questionnaire (Noy 2008, Khan, Keung et al. 2017, Khan, Keung et al. 2017). To perform the pilot evaluation of survey instrument, three external experts were requested. The requested experts include one from academics (Chongqing University, Chain) and two from industry (Virtual force-Pakistan and QSoft-Vietnam). The experts suggest some modifications related to the questionnaire structure and the questions for the collection of bibliographic data of survey participants. They further suggest putting the questions in a tabular form. We update the questionnaire by carefully considering the recommendation of experts, and the final used survey instrument is given in Appendix-B.

### 3.2.4 Data sources
The data sources play a vital role in targeting the potential population. The potential population is important to target as it is necessary for the collection of pure data. As the objective of this survey study was to get the insight of experts concerning to Continuous software engineering challenging factors identified from literature via SLR study. Though, to target the population, we used both professional Email addresses, Research-Gate, and LinkedIn. The snowballing technique was used to spread the survey questionnaire to the target geographically dispersed population

(Easterbrook, Singer et al. 2008, Finstad 2010, Khan, Keung et al. 2017). Snowball is an easy and cost-effective approach to collect the data from large and potential population.

The data was collectedfromDecember-2019 to March-2020. During the data collection process, a total of 102 responses were collected. All the received responses were checked, and nine responses were found uncompleted. By debating with the research team, we decided not to consider the uncompleted response for the data analysis process. Though, final 93 complete answers were considered for further analysis. The detail of respondents' bibliographic data are provided

### 3.2.3 Ethics approval
Once the survey questionnaire was finalized, we have conducted ethical approval for data collection from the "Research Ethics Board of College of Computer Science and Technology, Nanjing University of Aeronautics and Astronautics, Nanjing." After getting approval from the Research Ethical Board, we started the data collection process and made available the questionnaire for the targeted population.
in section-4.2.

### 3.2.5 Survey data analysis
We have adopted the frequency analysis method as it is an effective technique to analyze the quantitative and qualitative data. It is an appropriate approach to compared the respondent's opinions among the variables and group of variables (Bland 2015). The same approach has been considered by several researcher of other software engineering domains (Keshta, Niazi et al. 2017, Mahmood, Anwer et al. 2017, Akbar, Sang et al. 2018).

### 3.3 Phase 3: Fuzzy Set Theory and AHP
The implementation process of fuzzy AHP steps is discussed in this section.

### 3.3.1 Fuzzy set theory
The fuzzy set theory is an extend version of classical set theory developed by (Zadeh, Klir et al. 1996). That was considered to address the vagueness and uncertainties in the industry practices using multicriteria decision making problems. In the fuzzy set, a membership function µF(x) is characterized, which maps an object between 0 and 1. The definitions and preliminary of the fuzzy set theory are explained in subsequent sections:

Definition: "A triangular fuzzy number (TFN) F is denoted by a set (fl, fm, fu), as presented in Figure 4. The given equation"
(1) Defines the membership function µF(x) of F.

$$\mu_F(x) = \begin{cases} \dfrac{t - v^l}{v^m - v^l}, & v^l \leq t \leq v^m \\ \dfrac{v^u - t}{v^u - v^m}, & v^m \leq t \leq v^u \\ 0, & Otherwise \end{cases} \qquad (1)$$

"Where $v^l$, $v^m$ and $v^u$ are the crisp numbers denoting the lowest, most promising and highest possible values respectively".

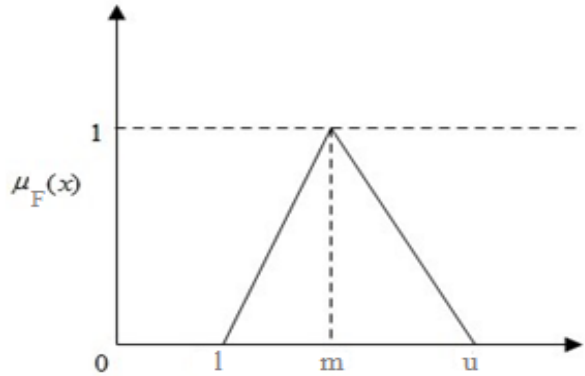

**Figure 4.** Triangular fuzzy number

The "algebraic operational laws using two TFNs, namely ($V_1$, $V_2$)" are given in Table 2.

**Table 2.** Triangular Fuzzy Numbers

| Operation Law | Expression |
| --- | --- |
| Addition ($V_1 \oplus V_2$) | $(v^l_1, v^m_1, v^u_1) \oplus (v^l_2, v^m_2, v^u_2) = (v^l_1 + v^l_2, v^m_1 + v^m_2, v^u_1 + v^u_2)$ |
| Subtraction ($V_1 \ominus V_2$) | $(v^l_1, v^m_1, v^u_1) \oplus (v^l_2, v^m_2, v^u_2) = (v^l_1 - v^l_2, v^m_1 - v^m_2, v^u_1 - v^u_2)$ |
| Multiplication ($V_1 \otimes V_2$) | $(v^l_1, v^m_1, v^u_1) \oplus (v^l_2, v^m_2, v^u_2) = (v^l_1 * v^l_2, v^m_1 * v^m_2, v^u_1 * v^u_2)$ |
| Division ($V_1 \oslash V_2$) | $(v^l_1, v^m_1, v^u_1) \oplus (v^l_2, v^m_2, v^u_2) = (v^l_1 / v^l_2, v^m_1 / v^m_2, v^u_1 / v^u_2)$ |
| Inverse ($V_1 \oplus V_2$) | $(v^l_1, v^m_1, v^u_1)^{-1} = (1/v^l_1, 1/v^m_1, 1/v^u_1)$ |
| For any real number k ($kV_1$) | $k(v^l_1, v^m_1, v^u_1) = k\,v^l_1, k\,v^m_1, k\,v^u_1$ |

### 3.3.2 Fuzzy AHP

The fuzzy AHP is a useful approach for "multicriteria decision making problems". The key benefit of fuzzy AHP is that it is easy to apply and understandable; and it can manage both quantitative and qualitative data. Following are the main steps adopted to perform the fuzzy AHP:

**Step1:** "Decompose the complex decision problem into the hierarchical structure" (Figure 5)
**Step2:** Determination of priority weights.
**Step3:** Apply the consistency check on each pairwise comparison matrix.
**Step4:** Determination of final ranking for each challenge and their respective categories" (Figure 5).

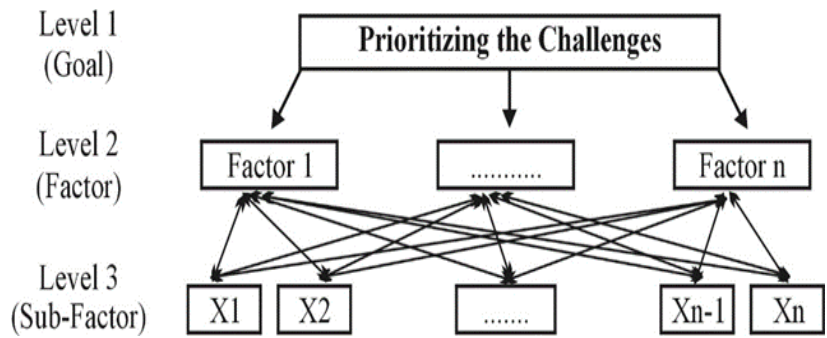

**Figure 5.** FAHP decision hierarchy

Hence, the classical AHP has some limitations due to the implementing the AHP in the Crisp environment, "Judgmental scale is unbalanced", and the "lack of ambiguity", "selection of judgment" are subjective. Though, fuzzy AHP is an updated version of AHP and that was develop to fix the uncertainties more effectively (Saaty

2013). The fuzzy AHP is effective to address the uncertainty and imprecise judgment of experts by handling the linguistic variables. The fuzzy AHP has been applied in various other domain (Sloane, Liberatore et al. 2002, Wenying 2009, Kabra, Ramesh et al. 2015, Yaghoobi 2018). In current study, we applied the fuzzy AHP that is introduced by (Chang 1996),which offer more consistent and accurate results compared with other approaches.

In a ranking problem, let X = {x1, x2,..., xn} indicate the factors of main categories as an object set and U = {u1, u2,..., un} indicates the factors a particular category as a goal set. By (Chang 1996)approach, every element is measured, and extent analysis for each goal (gi) is performed, respectively. Therefore, for each object, there are (m) extent analysis values that can be obtained with the following Equation (2) and (3):

$$V^1_{gi}, V^2_{gi}, ..., V^m_{gi}, \quad (2)$$

$$i = 1, 2, ..., n \quad (3)$$

Where, all $F^j_{gi}$, (j = 1, 2, ..., m) are triangular fuzzy numbers (TFNs).
The main steps of Chang's extent analysis approach (Chang 1996) are:

**Step 1:** The value of fuzzy synthetic extent with respect to the $^{ith}$ object can be defined using Eq. (4):

$$S_i = \sum_{j=1}^{m} V^j_{gi} \otimes \left[ \sum_{i=1}^{n} \sum_{j=1}^{m} V^j_{gi} \right]^{-1} \quad (4)$$

To achieve the expression $\sum_{j=1}^{m} V^j_{gi}$, evaluate the fuzzy addition operation extent analysis such as:

$$\sum_{j=1}^{m} V^j_{gi} = (\sum_{j=1}^{m} v^l_{gi}, \sum_{j=1}^{m} v^m_{gi}, \sum_{j=1}^{m} v^u_{gi}) \quad (5)$$

and to achieve the expression $\left[ \sum_{i=1}^{n} \sum_{j=1}^{m} V^j_{gi} \right]^{-1}$, the fuzzy addition operation is executed on $V^j_{gi} (j = 1, 2, .....m)$ value, as follow:

$$\sum_{i=1}^{n} \sum_{j=1}^{m} V^j_{gi} = (\sum_{i=1}^{n} v^l_i, \sum_{i=1}^{n} v^m_i, \sum_{i=1}^{n} v^u_i) \quad (6)$$

and finally, calculate the inverse of the vector with the help of Eq. (7):

$$\left[ \sum_{i=1}^{n} \sum_{j=1}^{m} V^j_{gi} \right]^{-1} = (\frac{1}{\sum_{i=1}^{n} v^l_i}, \frac{1}{\sum_{i=1}^{n} v^m_i}, \frac{1}{\sum_{i=1}^{n} v^u_i}) \quad (7)$$

**Step 2:** As $F_a$ and $F_b$ are two fuzzy triangular numbers, then the degree of possibility of $V_a = (v^l_a, v^m_a, v^u_a) \geq V_b = (v^l_b, v^m_b, v^u_b)$ is defined as follows and the Eq. 8 can also be similarly specified as below:

$$V(V_a \geq V_b) = sup[min(\mu_{va}(x), (\mu_{vb}(x))] \quad (8)$$

$$V(V_a \geq V_b) = hgt(V_a \cap V_b) = \mu_{v_a}(d) = \begin{cases} 1 & \text{if } v^m_a \geq v^m_b \\ \dfrac{v^u_a - v^l_b}{(v^u_a - v^m_a) + (v^m_b - v^l_b)} & v^l_b \leq v^u_a \\ 0 & \text{Otherwise} \end{cases} \quad (9)$$

Here, d indicates the ordinate of the highest intersection point between D, $\mu_{Va}$ and $\mu_{Vb}$ (Figure 6). The values of $T_1(V_a \geq V_b)$ and $T_2(V_a \geq V_b)$ are required for determining the value of $P_1$ and $P_2$.

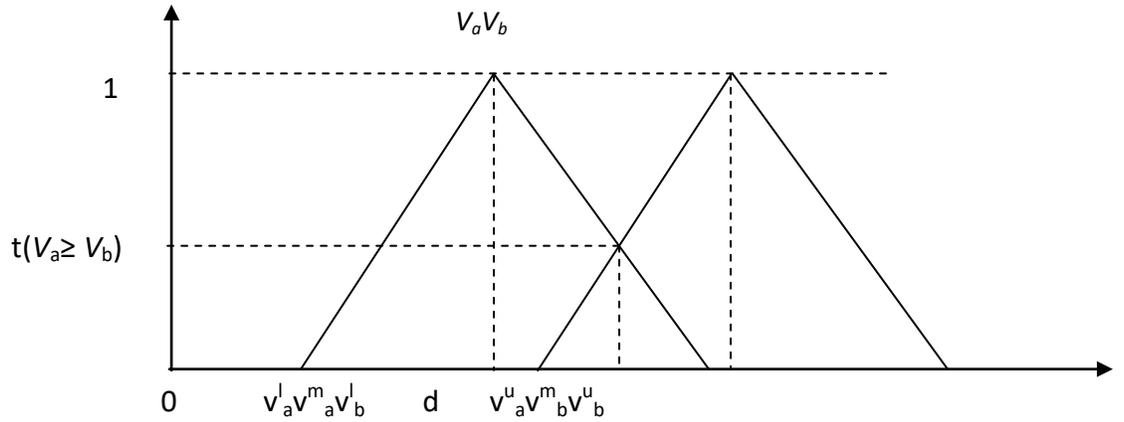

**Figure 6.** Triangular Fuzzy number

**Step 3:** Calculate the overall degree of possibility of a convex fuzzy number, and the other convex fuzzy numbers $V_i$ ($i = 1, 2,..., k$) can be descripted as follow.

$$T(V \geq V_1, V_2, V_3 .... V_k) = \min T(V \geq V_i) \quad (10)$$

Assuming that,

$$d'(V_i) = \min T(V_i \geq V_k) \quad (11)$$

for k = 1,2,...,n; $k \neq i$.

With the help of Eq. 12, determine the weight vector using Eq. 11.

$$W' = (d'(V_1), d'(V_2), d'(V_3), ..... d'(V_n)) \quad (12)$$

Where, $V_i$ ($i = 1,2,...,n$) are $n$ distinct elements.

**Step 4:** The normalization, the normalized weight vectors are in equation 13, and the result will be a non-fuzzy number which renders the priority weight of the challenge:

$$W = (d(V_1), d(V_2), d(V_3), \ldots d(V_n)) \qquad (13)$$

Where $W$ is a non-fuzzy number.

**Step 5: Checking consistency ratio**: The developed pairwise comparison matrixes should be consistent in fuzzy AHP analysis (Shameem, Kumar et al. 2018). Hence, it is mandatory to measure the consistency ratio of the pairwise comparison matrixes. To address this, the "graded mean integration" method is used for defuzzifying the matrix. A triangular fuzzy number, denoted as P = (l, m, u), can be defuzzified to a crisp number as follows:

$$P_{crisp} = \frac{(4m + l + u)}{6} \qquad (14)$$

Besides the defuzzification of every element of matrix, the consistency ration (CR) of each pairwise comparison matrix is easy to determine aiming to check as to determine the value of CR is less than 0.10 or not. For this, two primary parameters, i.e., "consistency index" (CI) and "consistency ratio" (CR) are considers and both are defined in Equations 14 and 15, respectively.

$$CI = \frac{I_{max} - n}{n - 1} \qquad (15)$$

$$CR = \frac{CI}{RI} \qquad (16)$$

Where,
$I_{max}$: presents the largest eigenvalue of the pairwise matrix.
n: presents the number of values being compared.
RI: "the random index and its value can opt from Table 3".
CR: If the value of CR is less than 0.1, then it denotes the consistent pairwise comparison matrix.

Table 3: **"Random consistency index (RI) with respect to matrix size"**

| "Matrix size" | 1 | 2 | 3 | 4 | 5 | 6 | 7 | 8 | 9 | 10 |
|---|---|---|---|---|---|---|---|---|---|---|
| RI | 0 | 0 | 0.58 | 0.9 | 1.12 | 1.24 | 1.32 | 1.41 | 1.45 | 1.49 |

**4. Results and discussion**
This section contains the results and analysis of this study.

**4.1 Findings of SLR study**
The phases of the SLR approach were carefully executed to extract the challenging critical factors of Continuous software engineering practices. The list of investigated 22 challenges were enlisted in Table 4.

Table 4: List of challenges

| Sr. No | Challenges | IDs (N=78) |
|---|---|---|
| C1 | Moving from legacy infrastructure to microservices | [PS2], [PS9], [PS14], [PS20], [PS51], [PS54], [PS56], [PS57], [PS59], [PS61], [PS63], [PS64], [PS65], [PS68] |
| C2 | No Continuous software engineering center of excellence | [PS11], [PS20], [PS21], [PS23], [PS27], [PS31], [PS35], [PS40], [PS44], [PS61], [PS62], [PS75] |
| C3 | Lack of Continuous software engineering metrics | [PS4], [PS7], [PS9], [PS13], [PS15], [PS18], [PS22], [PS23], [PS30] [PS40], [PS51], [PS54], [PS56] |
| C4 | Lack of service virtualization | [PS1], [PS7], [PS10], [PS13], [PS15], [PS19], [PS24], [PS29], [PS30], [PS63], [PS75], [PS76], [PS78] |
| C5 | Building and maintaining the deployment pipeline | [PS2], [PS3], [PS5], [PS9], [PS10], [PS12], [PS13], [PS17], [PS18], [PS21], [PS22], [PS49], [PS51], [PS55], [PS56], [PS61], [PS62], [PS63], [PS66] [PS67], [PS68], [PS70], [PS74] |
| C6 | Overcoming the Dev versus Ops mentality | [PS8], [PS11], [PS20], [PS21], [PS24], [PS26], [PS29], [PS31], [PS35], [PS40], [PS46], [PS47], [PS48], [PS53], [PS61], [PS69], |
| C7 | Lack of integrated tools architecture | [PS5], [PS10], [PS20], [PS34], [PS35], [PS36], [PS37], [PS40], [PS44], [PS67], [PS68], [PS70] |
| C8 | Continuous software engineering and regulatory compliance | [PS5], [PS7], [PS13], [PS15], [PS19], [PS25], [PS27], [PS29], [PS30], [PS32], [PS33], [PS39], [PS40], [PS46], [PS47], [PS66], [PS71], [PS73], [PS75] |
| C9 | Traceability across the Continuous software engineering landscape | [PS2], [PS9], [PS13], [PS17], [PS21], [PS23], [PS29], [PS30], [PS32], [PS41], [PS45], [PS47], [PS48], [PS53], [PS58], [PS59], [PS61], [PS64], [PS69], [PS78] |
| C10 | Artifact management issues | [PS9], [PS11], [PS12], [PS14], [PS16], [PS18], [PS21], [PS22], [PS25], [PS27], [PS29], [PS34], [PS35], [PS39], [PS42], [PS43], [PS45], [PS46] [PS67], [PS68], [PS69], [PS74], [PS75], [PS76], [PS78] |
| C11 | Configuration Management | [PS3], [PS10], [PS24], [PS29], [PS31], [PS33], [PS34], [PS36], [PS39], [PS40], [PS50], [PS53], [PS56], [PS61], [PS62], [PS72], [PS75], [PS76] |
| C12 | Resources accountability issues | [PS9], [PS15], [PS19], [PS25], [PS29], [PS33], [PS34], [PS36], [PS37], [PS39], [PS41], [PS54], [PS57], [PS60], [PS62], [PS67], [PS78] |
| C13 | Incident handling issues | [PS10], [PS11], [PS21], [PS28], [PS34], [PS36], [PS38], [PS39], [PS40], [PS42], [PS43], [PS47], [PS54], [PS55], [PS59], [PS66], [PS76] |
| C14 | Resistance to change | [PS4], [PS19], [PS20], [PS22], [PS27], [PS30], [PS32], [PS34], [PS37], [PS38], [PS48], [PS49], [PS50], [PS53], [PS56], [PS57], [PS58], [PS60], [PS62], [PS75] |
| C15 | Lack of trust relationship | [PS8], [PS18], [PS26], [PS29], [PS34], [PS35], [PS37], [PS39], [PS40], [PS41], [PS43], [PS47], [PS50], [PS55], [PS57], [PS62], [PS69], |
| C16 | Communication and Collaboration issues | [PS6], [PS16], [PS19], [PS53], [PS55], [PS56], [PS58], [PS60], [PS62], [PS63], [PS64], [PS69], [PS71], [PS76] |
| C17 | Disintermediation of roles within teams | [PS4], [PS7], [PS11], [PS13], [PS16], [PS21], [PS22], [PS30], [PS33], [PS36], [PS41], [PS43], [PS45], |
| C18 | Inconsistent environments | [PS11], [PS26], [PS29], [PS30], [PS33] [PS37], [PS43], [PS44], [PS49],[PS54], [PS57], [PS58], [PS63] |
| C19 | Lack of feedback and bugs prioritization | [PS7], [PS15], [PS16], [PS19], [PS20], [PS23] [PS27], [PS30], [PS31], [PS34],[PS43], [PS45], [PS53], [PS57], [PS68] |
| C20 | Lack of flexibility due to rigid Industrial constraints | [PS2], [PS3], [PS6], [PS16], [PS17], [PS20], [PS26], [PS29], [PS31], [PS36], [PS38], [PS42], [PS46], [PS48], [PS50], [PS57], [PS59], [PS65], [PS67], [PS69], [PS71], [PS73], [PS74] |
| C21 | Lack of strategic suggestions from leadership | [PS2], [PS3], [PS11], [PS17], [PS27], [PS37], [PS39], [PS42], [PS44], [PS49], [PS43], [PS50], [PS51], [PS57], [PS58], [PS60], [PS75], [PS76], [PS77], [PS78] |
| C22 | Heterogeneity in development and operational structure | [PS2], [PS11], [PS16], [PS17], [PS20], [PS22] [PS27], [PS29], [PS31], [PS34],[PS38], [PS40], [PS43], [PS47], [PS48], [PS50], [PS56], [PS62], [PS73], |

The identified challenges were further mapped in the core phases of the CAMS model, developed by Edwards and Willis(Guthrie 2019). The critical aspects of CAMS include "Culture", "Automation", "Measurement", and "Sharing". The CAMS model consists of a set of variables considered by various practitioners for the successful implementation of Continuous software engineering practices in the software industry. We develop a mapping team consists of three authors of this study (Author number 1, 3, 4). All the participants of the mapping team were continuously involved and using the critical steps of the coding scheme ((i.e., "code," "sub-categories," "categories and theory\framework"), all the challenging factors were mapped and developed a framework as given in Figure 7. The principal objective of mapping is to perform the fuzzy AHP analysis.

**Culture:** "Culture is defined by the interaction of people and groups and is driven by behavior. Substantial communication improvement can result when there is a mutual understanding of others and their goals and responsibilities".

**Automation:** "Automation can save time, effort, and money, just like culture, it truly focuses on people and processes and not just tools. The impact of implementing infrastructure as code as well as using continuous integration and continuous delivery pipelines can be magnified after understanding an organization's culture and goals. It helps to think of automation as an accelerator that enhances the benefits of Continuous software engineering as a whole".

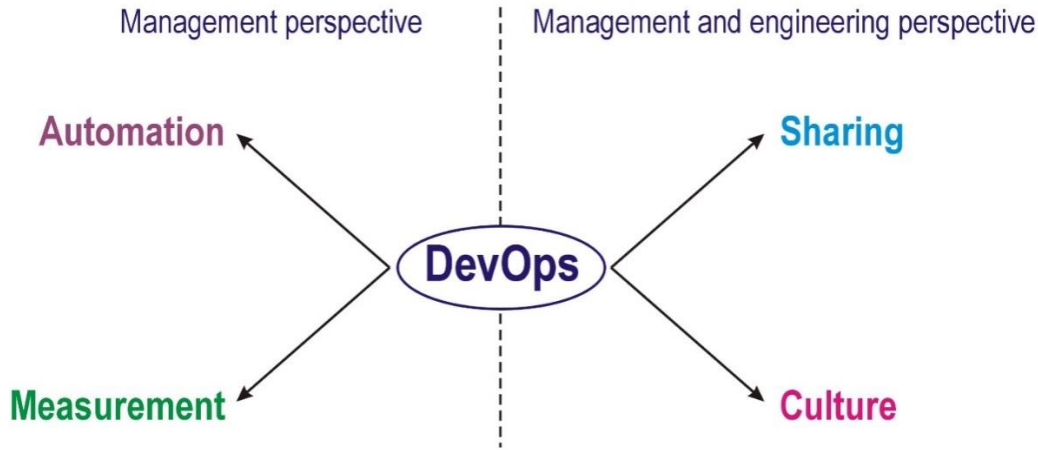

Figure 7: conceptual modeling of Continuous software engineering parameters

**Measurement:** "The measurement is helpful to determine the progresses and its intended direction. There are two main bumps that might be occur while using matrices i.e. (1) to make sure the parameters are correct ones and (2) to incentivize the right metrics. The Continuous software engineering encourage to see the forest from the trees by viewing the entire operation and evaluating it as a whole and not just focusing on small parts. Primary metrics include (but are certainly not limited to) income, costs, revenue, mean time to recovery, mean time between failures, and employee satisfaction".

**Sharing:** "Continuous software engineering processes, similar to agile and scrum, place a very high premium on transparency and openness. Spreading knowledge helps to tighten feedback loops and enables the organization to improve continuously. This collective intelligence makes the team a more efficient unit and allows it to become greater than just the sum of its parts."

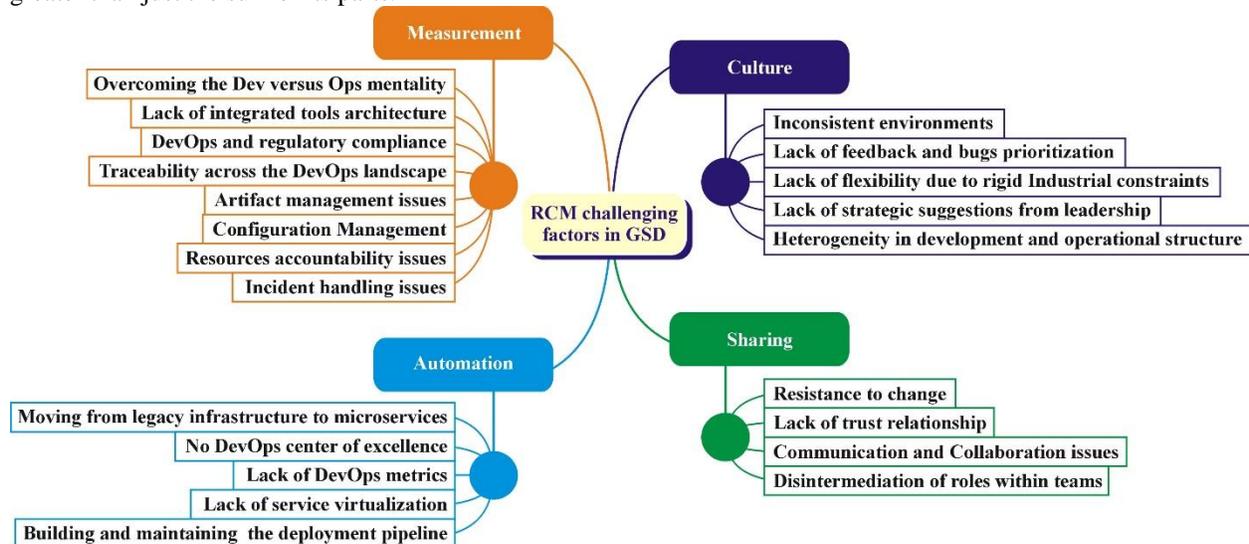

Figure 8: Mapping of investigated challenges into CAMS

### 4.2 Empirical study results
To verify the finding of the SLR study, the questionnaire survey study was conducted with experts and the analyzed responses of survey participates are given in subsequent sections.

### 4.2.1 Demographic data analysis of survey participants
The detail demographic information of the survey participants was collected during the data collection process. (Patten 2016), stated that "the demographic data provides information about survey respondents and is essential for the determination of whether the participants in a particular study are a representative sample of the target population for results generalization purposes or not."(Finstad 2010) underlined that the bibliographic data of survey participants give the insight of survey respondents, which shows the maturity level of the collected data set. (Altman, Machin et al. 2013), underlined that the information of survey participants assists in determining "what your target population is and what they are thinking about." Though, by seeking the importance of respondents' bibliographic data, we have analyses the respondent's data concerning to organization size, respondents designation and organization size. The brief analysis is discussed in the following sections.

**4.2.1.1 Respondent's designation**
(Finstad 2010) underlined the implication and priority of the influencing factors that vary regarding the designation of the respondents. Furthermore,(Niazi, Mahmood et al. 2016)defines that the impact of a factor depending upon the position of the practitioner, and they further stated that the influence of an element could be ranked exactly if the respondent frequently experiences to deal with that factor. The responder's designation based analysis is presented in Figure 9 that shows the most of the survey respondents are project manager. According to the results the most common respondents' designations are: "project manager", "software developer", "researcher" (Figure 9).

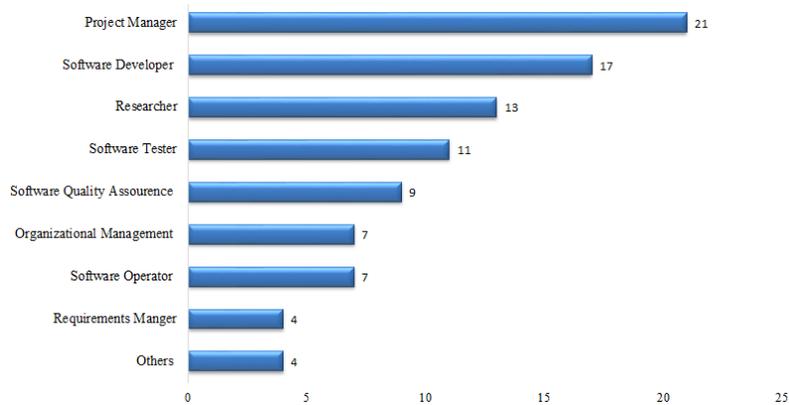
Figure 9: Analysis of the survey participant's designations

**4.2.1.2 Respondent's experience**
The experience of the survey participants reported in the questionnaire was also analyzed. The mean and medium were calculated, and the results show 7.5 and 5, respectively; this renders the young pool of survey participants. Besides, we also observed the significant variations in the experience of the survey participants. The detailed results of the survey participates are graphically shown in Figure 10.

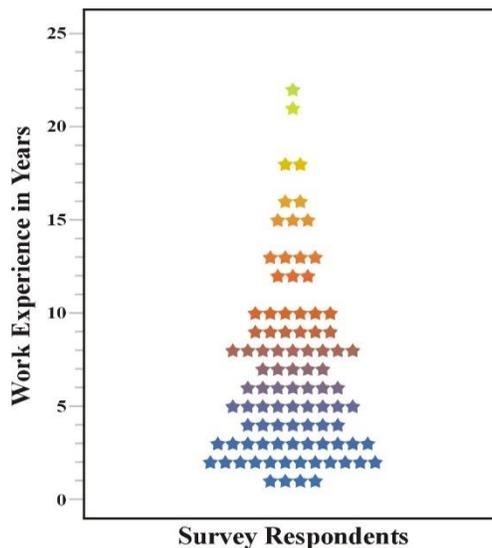
Figure 10: Experience of survey respondents

### 4.2.1.3 Organization size

The respondent's bibliographic data were also concerning to their organizational size. The organizations were classified on small, medium, and large scale with respect to the definition of Australian bureau of statistics (Trewin 2002), i.e. "(SMALL, 0–19 employees), (MEDIUM, 20–200 employees), and (LARGE, ≥200 employees)"(Trewin 2002). (Akbar, Sang et al. 2019) indicated that the organization size is also a critical entity to assess the maturity level and explore of survey participants. The results presented in Figure 11 renders that 31(33%), 37 (40%), and 25 (27%)respondents are from a small, medium and large scale of firms, respectively. The detail of organization size based analysis is presented in Figure 11.

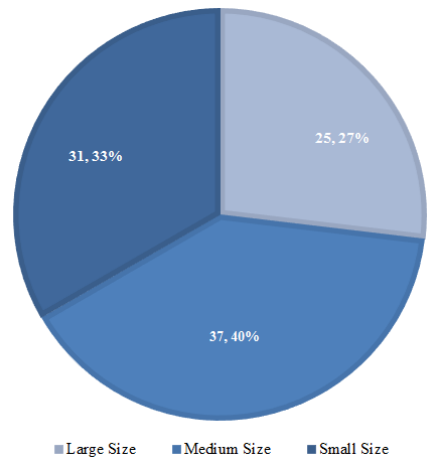

Figure 11: Participants organizations size

### 4.2.2 Responses against Continuous software engineering challenges (RQ3)

The basic objective of empirical study was to get insight into the industry practitioners concerning the Continuous software engineering challenges identified via SLR study. The responses collected against the Continuous software engineering challenging factors were mainly categorized as: positive ("agree, strongly agree"), negative ("disagree, strongly disagree"), and "neutral". The positive category presents the frequency of those survey respondents who are considered the identified challenging factors that could negatively influence the Continuous software engineering practices. The negative group presents the frequency of those respondents who do not agree with the identification of SLR study. The neutral category shows the frequency of survey participates who are not sure about the effect of identified factors concerning Continuous software engineering activities. The detail results are given in Table 5.

Table 5: Empirical investigation

| | | Number of Responses (N=93) | | | | | | | |
| | | Positive | | | Negative | | | Neutral | |
| S.NO | List of challenges | S.A | A | % | D | S.D | % | N | % |
|---|---|---|---|---|---|---|---|---|---|
| **P1** | **Automation** | **31** | **52** | **89** | **0** | **5** | **5** | **5** | **5** |
| C1 | Overcoming the Dev versus Ops mentality | 26 | 43 | 74 | 4 | 8 | 13 | 12 | 13 |
| C2 | Moving from legacy infrastructure to microservices | 24 | 51 | 81 | 1 | 6 | 8 | 11 | 12 |
| C3 | Resistance to change | 29 | 38 | 72 | 3 | 9 | 13 | 14 | 15 |
| C4 | Lack of integrated tools architecture | 24 | 50 | 80 | 2 | 8 | 11 | 9 | 10 |
| C5 | No Continuous software engineering center of excellence | 27 | 40 | 72 | 3 | 9 | 13 | 14 | 15 |
| **P2** | **Measurement** | **29** | **48** | **83** | **2** | **6** | **9** | **8** | **9** |
| C6 | Lack of Continuous software engineering metrics | 31 | 39 | 75 | 4 | 7 | 12 | 12 | 13 |
| C7 | Continuous software engineering and regulatory compliance | 26 | 38 | 69 | 6 | 7 | 14 | 16 | 17 |
| C8 | Lack of service virtualization | 23 | 40 | 68 | 4 | 11 | 16 | 15 | 16 |
| C9 | Traceability across the Continuous software engineering landscape | 30 | 49 | 85 | 0 | 7 | 8 | 7 | 8 |
| C10 | Inconsistent environments | 31 | 40 | 76 | 4 | 8 | 13 | 10 | 11 |
| C11 | Building and maintaining the deployment pipeline | 27 | 38 | 70 | 5 | 7 | 13 | 16 | 17 |

| C12 | Incident handling issues | 29 | 40 | 74 | 4 | 6 | 11 | 14 | 15 |
| C13 | Artifact management issues | 30 | 40 | 75 | 3 | 8 | 12 | 12 | 13 |
| **P3** | **Sharing** | **24** | **56** | **86** | **3** | **3** | **6** | **7** | **8** |
| C14 | Lack of trust relationship | 30 | 38 | 73 | 5 | 6 | 12 | 14 | 15 |
| C15 | Configuration Management | 26 | 44 | 75 | 3 | 7 | 11 | 13 | 14 |
| C16 | Communication and Collaboration issues | 39 | 34 | 78 | 2 | 4 | 6 | 14 | 15 |
| C17 | Resistance to adopt Continuous software engineering | 27 | 40 | 72 | 5 | 6 | 12 | 15 | 16 |
| **P4** | **Culture** | **37** | **49** | **92** | **0** | **2** | **2** | **5** | **5** |
| C18 | Lack of feedback and bugs prioritization | 30 | 38 | 73 | 5 | 8 | 14 | 12 | 13 |
| C19 | Resources accountability issues | 27 | 43 | 75 | 6 | 7 | 14 | 10 | 11 |
| C20 | "Lack of flexibility due to rigid Industrial constraints" | 33 | 46 | 85 | 4 | 5 | 10 | 5 | 5 |
| C21 | "Lack of strategic suggestions from leadership" | 27 | 39 | 71 | 5 | 8 | 14 | 14 | 15 |
| C22 | "Heterogeneity in development and operational structure" | 29 | 44 | 78 | 0 | 9 | 10 | 11 | 12 |

The concluded results are given in Table 5, which renders that majority of the survey participant's agree with the identified challenges as they have a negative relation with Continuous software engineering related to real-world practices. The frequency analysis shows that all the challenging factors considered ≥70% of the survey participants, instead of two challenges, i.e., C7 (Continuous software engineering and regulatory compliance, 69%) and C8 (Lack of service virtualization, 68%). We further noted that C20 ("Lack of flexibility due to rigid industrial constraints", 85%) was the highest reported challenging factors by the survey respondents.

We observed that P4 (Culture, 92%) was the highest considered category of the investigated challenging factors. P1 (Automation, 89%) and P3 (Sharing, 86%) were considered as the second and third most significant important categories of challenges.

The negative category shows C8 (Lack of service virtualization, 16%) is the highest-ranked challenge factors, this renders that 16% of the respondents are not agree with the C8 as a challenging factor for Continuous software engineering practices. C7 (Continuous software engineering and regulatory compliance, 14%), C18 ("Lack of feedback and bugs prioritization", 14%), C19 (Resources accountability issues, 14%) and C21 ("Lack of strategic suggestions from leadership", 14%) are mention as the second highest ranked challenging factors.

We further observed that C11 (Building and maintaining the deployment pipeline, 17%), C8 (Lack of service virtualization, 16%), C17 (Resistance to adopt Continuous software engineering, 16) are declared as the first and second highest ranked challenges for Continuous software engineering paradigm in software organizations, respectively.

**4.3 Application of fuzzy AHP**
This section contains the fuzzy-AHP analysis of the explored challenges and their categories. The priority of the challenges were determined using the step by step protocols of fuzzy AHP, as presented in above (section 3.4).

**Step-1: (categorize the complicated problems into hierarchy structure)**
To perform the fuzzy AHP analysis, the complicated problem is divided to an interconnected decision making elements. (Shameem, Kumar et al. 2018), (Albayrak and Erensal 2004). The complicated problem is classified at minimum of 3 stages as presented in Figure 5, whereas the key aim of the problems is indicated at top level, the categories of challenge and their corresponding challenges are presented at stage 2 and 3, respectively. The proposed hierarchy structure is presented in Figure 12.

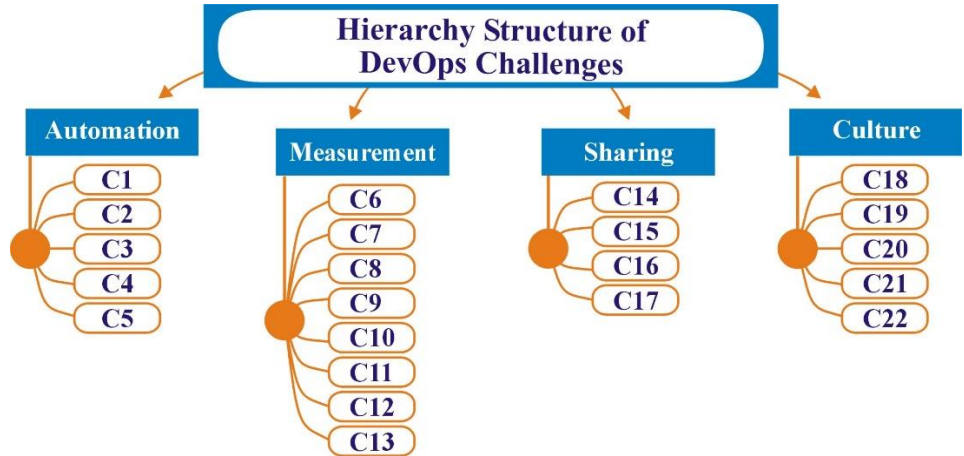

Figure 12: Proposed hierarchal structure of the problem

**Step-2: Pairwise comparison**
The pairwise comparison was conducted based on the opinions of experts. To conduct the pairwise comparison, we have develop a questionnaire and contacted with the experts involved in first survey. The developed fuzzy-AHP questionnaire (Appendix-D) was sent to the experts and the 29 responses were collected. The collected complete responses were further manually analysed to find the incomplete entries. However, we did not found any incomplete entries and the collected 29 responses were considered for further analysis. In order to generalize results of fuzzy-AHP analysis the data of 29 respondents might not strong enough. As the fuzzy AHP is a subjective methodology and the small data size is acceptable (Sloane, Liberatore et al. 2002, Wen-ying 2009, Kabra, Ramesh et al. 2015, Yaghoobi 2018).

It is noted that the existing studies used the small of data size for fuzzy AHP analysis. In example, (Akbar, Shameem et al.)performed the fuzzy AHP analysis considering the data collected form 23 experts. (Shameem, Kumar et al. 2018)used the opinion of 5 experts for perfume AHP analysis. (Cheng and Li 2002)conducted pairwise comparison considering the opinions of 9 experts. Moreover, we found that (Wong and Li 2008) conducted the fuzzy AHP analysis using the data collected from 9 experts. Though, considering data size of existing studies, we are confident that the data collected from 29 experts is justified for generalizing the result of fuzzy AHP study. The collected opinions were converted into geometric mean aiming to develop the pairwise comparison matrixes.

The geometric mean is useful method to transform the judgement of survey respondents into TFN numbers. In this study, we have used the following formula of geometric mean:

$$\text{Geometric mean} = \sqrt[n]{r_1 \times r_2 \times r_3 \ldots \ldots r_n}$$

r="Weight of each response"
n="Number of responses"

Table 6: "Triangular fuzzy conversion scale (Bozbura, Beskese et al. 2007)".

| "Linguistic scale" | "Triangular fuzzy scale" | "Triangular fuzzy reciprocal scale" |
|---|---|---|
| "Just equal (JE)" | (1, 1, 1) | (1, 1, 1) |
| "Equally important (EI)" | (0.5, 1, 1.5) | (0.6, 1, 2) |
| "Weakly important (WI)" | (1, 1.5, 2) | (0.5, 0.6, 1) |
| "Strongly more important (SMI)" | (1.5, 2, 2.5) | (0.4, 0.5, 0.6) |
| "Very strongly more important (VSMI)" | (2, 0.5, 3) | (0.3, 0.4, 0.5) |
| "Absolutely more important (AMI)" | (2.5, 3, 3.5) | (0.2, 0.3, 0.4) |

**Step-3: Test the consistency of the pair-wise matrix**
The step by step procedure to calculate the consistency are presented in this section. The pairwise comparison matrixes should be consistent in fuzzy AHP. To do this, the Likert scale categories (Table 6) are considered. A triangular fuzzy number of the pair-wise comparison matrix of the main categories are defuzzified to crisp number using Equation 14 and obtained the corresponding Fuzzy Crisp Matrix (FCM) as shown in Table 7:

Table 7: "Fuzzy-Crisp Matrix (FCM) for challenges categories"

|            | Automation | Measurement | Sharing | Culture |
|------------|------------|-------------|---------|---------|
| Automation | 1.00       | 2.50        | 1.50    | 2.00    |
| Measurement| 0.40       | 1.00        | 0.40    | 0.73    |
| Sharing    | 0.65       | 2.50        | 1.00    | 1.50    |
| Culture    | 0.50       | 1.43        | 0.65    | 1.00    |
| Column Sum | 2.55       | 7.43        | 3.55    | 5.23    |

The largest Eigenvector ($I_{max}$) value of the FCM matrix is determined by taking the sum of the elements of each column of FCM matrix (Table 7). The determined sum is further divided with each element of FCM matrix and take the average of each element of each row, as presented in Table 8.

Table 8: Normalized matrix of challenges categories

|            | Automation | Measurement | Sharing | Culture | Priority vector weight |
|------------|------------|-------------|---------|---------|------------------------|
| Automation | 0.39217    | 0.33633     | 0.42255 | 0.38215 | 0.38431                |
| Measurement| 0.15685    | 0.13454     | 0.11267 | 0.14014 | 0.13617                |
| Sharing    | 0.25491    | 0.33635     | 0.28168 | 0.28664 | 0.28986                |
| Culture    | 0.19607    | 0.19284     | 0.18311 | 0.19109 | 0.19077                |

$I_{max} = \Sigma ([\Sigma C_j] \times \{W\})$                                            (17)
Where, $\Sigma C_j$ = sum of the columns of Matrix [C] (Table 8),
W= weight vector (Table 8), hence,
$I_{max}$ = 2.55*0.38431 + 7.43*0.13617 + 3.55*0.28986 + 5.23*0.19077 = 4.0162

Thus, considering the value of $I_{max}$ (4.0162), the dimension of FCM is 4. Random Consistency Index (RI) for n=4 is 0.9 (Table 3). Using the equation 15 and 16, we calculated the consistency ration for each pairwise matrix as:

$$CI = \frac{I_{max} - n}{n - 1} = \frac{4.0162 - 4}{4 - 1} = 0.0053865$$

$$CR = \frac{CI}{RI} = \frac{0.0053865}{0.9} = 0.0059850$$

The determined value of CR is 0.005985<0.10; hence, the pairwise comparison matrix of challenges categories is consistence and acceptable for fuzzy AHP. By considering the same steps, the consistency ratio for all the challenging factors categories were determined and the results are presented in Table 10, 11, 12, 13, and 14, respectively.

**Step 4: Determining the local priority weight of each challenge**
The local ranking presents the priority order of a challenge in their own category. The local ranking assists the practitioners to address the Continuous software engineering challenges mentioned in a particular category. To do this, the local weigh of each challenge was determined using the Equation 3.

Firstly, the synthetic extent values of categories (C1, C2, C3, C4) were calculated, and the priority weight of the attributes was mention using Equation 4. In this section, we exemplary present the weight calculation of main categories of the challenges.

$$\sum_{i}^{n}\sum_{j}^{m} V_{gi}^{j} = (1,1,1) + (2, 2.5, 3) + .... + (0.5, 0.6, 1) + (1,1,1) = (14.6, 18.5, 23.6)$$

$$\left[\sum_{i}^{n}\sum_{j}^{m}V_{gi}^{j}\right]^{-1} = (\frac{1}{23.6},\frac{1}{18.6},\frac{1}{14.6}) = (0.042373, 0.054054, 0.068493)$$

$$\sum_{j=1}^{m}V_{g1}^{j} = (1,1,1)+(2,2.5,3)+(1,1.5,2)+(1.5,2,2.5) = (5.5,7.0,8.5)$$

$$\sum_{j=1}^{m}V_{g2}^{j} = (0.3,0.4,0.5)+(1,1,1)+(0.3,0.4,0.5)+(0.5,1,1.5) = (2.1,2.8,3.5)$$

$$\sum_{j=1}^{m}V_{g3}^{j} = (0.5,0.6,1)+(2,2.5,3)+(1,1,1)+(1,1.5,2) = (4.5,5.6,7.0)$$

$$\sum_{j=1}^{m}V_{g4}^{j} = (0.4,0.5,0.6)+(0.6,1,2)+(0.5,0.6,1)+(1,1,1) = (2.5,3.1,4.6)$$

The C1, C2, C3, and C4 represent the synthesis values of four challenges which were calculated using Equation 4 as follow:

$$CF1 = \sum_{j}^{m}V_{g1}^{j} \otimes \left[\sum_{i}^{n}\sum_{j}^{m}V_{gi}^{j}\right]^{-1}$$

$$= (5.5,7.0,8.5) \otimes (0.042373, 0.054054, 0.068493) = (0.233051, 0.378378, 0.582192)$$

$$CF2 = (2.1,2.8,3.5) \otimes (0.042373, 0.054054, 0.068493) = (0.088983, 0.151351, 0.239726)$$

$$CF3 = (4.5,5.6,7.0) \otimes (0.042373, 0.054054, 0.068493) = (0.190678, 0.302703, 0.479452)$$

$$CF4 = (2.5,3.1,4.6) \otimes (0.042373, 0.054054, 0.068493) = (0.105932, 0.167568, 0.315068)$$

The degree of possibility using Equation 9, as presented in Table 9, is determined.

Table 9. V values and d values for each category

|  | Automation | Measurement | Sharing | Culture | d (Priority Weight) |
|---|---|---|---|---|---|
| V (Automation ≥….) | 1 | 1 | 1 | 1 | 1 |
| V (Measurement ≥….) | 0.028563 | 1 | 0.24475 | 0.89191 | 0.02854 |
| V (Sharing ≥….) | 0.76504 | 1 | 1 | 0.67605 | 0.76506 |
| V (Culture ≥….) | 0.28007 | 1 | 0.47933 | 1 | 0.28007 |

In Table 9, the last column indicates the degree of possibility for each category that is determined using Equation 11 by considering the minimum value of each row.

Hence, the calculated weight vector is $W' = (1, 0.028563, 0.76504, 0.28009)$.

Once, the value weight vector were normalized, the significance of attributes were determine i.e. W = (0.482233, 0.013764, 0.368925, 0.135066). The determined results shows that 'Culture' is the most significant category or the challenging factors.

Table 10: Pairwise comparison of Automation category

| | P1 (Automaton) | | | | |
|---|---|---|---|---|---|
| | C1 | C2 | C3 | C4 | C5 |
| C1 | (1,1,1) | (0.3, 0.4, 0.5) | (0.4, 0.5, 0.6) | (1.5, 2, 2.5) | (0.4, 0.5, 0.6) |
| C2 | (2, 2.5, 3) | (1,1,1) | (2, 2.5, 3) | (0.5, 1, 1.5) | (1, 1.5, 2) |
| C3 | (1.5, 2, 2.5) | (0.3, 0.4, 0.5) | (1,1,1) | (2, 2.5, 3) | (2.5, 3, 3.5) |
| C4 | (0.4, 0.5, 0.6) | (0.6, 1, 2) | (0.3, 0.4, 0.5) | (1,1,1) | (0.5, 0.6, 1) |
| C5 | (1.5, 2, 2.5) | (0.5, 0.6, 1) | (0.2, 0.3, 0.4) | (1, 1.5, 2) | (1,1,1) |

$I_{max} = 5.51$, CI = 0.13, CR = 0.93

Table 11: Pairwise comparison of Measurement category

| | P2 (Measurement) | | | | | | | |
|---|---|---|---|---|---|---|---|---|
| | C6 | C7 | C8 | C9 | C10 | C11 | C12 | C13 |
| C6 | (1,1,1) | (1, 1.5, 2) | (2.5, 3, 3.5) | (0.6, 1, 2) | (1.5, 2, 2.5) | (1, 1.5, 2) | (0.5, 0.6, 1) | (0.3, 0.4, 0.5) |
| C7 | (0.5, 0.6, 1) | (1,1,1) | (0.5, 0.6, 1) | (1, 1.5, 2) | (0.4, 0.5, 0.6) | (1, 1.5, 2) | (2, 2.5, 3) | (1, 1.5, 2) |
| C8 | (0.2, 0.3, 0.4) | (1, 1.5, 2) | (1,1,1) | (0.5, 1, 1.5) | (0.5, 0.6, 1) | (0.5, 0.6, 1) | (1, 1.5, 2) | (0.5, 0.6, 1) |
| C9 | (0.5, 1, 1.5) | (0.5, 0.6, 1) | (0.6, 1, 2) | (1,1,1) | (0.2, 0.3, 0.4) | (2, 2.5, 3) | (0.5, 1, 1.5) | (2, 2.5, 3) |
| C10 | (0.4, 0.5, 0.6) | (1.5, 2, 2.5) | (1, 1.5, 2) | (2.5, 3, 3.5) | (1,1,1) | (0.4, 0.5, 0.6) | (0.2, 0.3, 0.4) | (1, 1.5, 2) |
| C11 | (0.5, 0.6, 1) | (0.5, 0.6, 1) | (1, 1.5, 2) | (0.3, 0.4, 0.5) | (1.5, 2, 2.5) | (1,1,1) | (0.4, 0.5, 0.6) | (2, 2.5, 3) |
| C12 | (1, 1.5, 2) | (0.3, 0.4,0.5) | (0.5, 0.6, 1) | (0.6, 1, 2) | (2.5, 3, 3.5) | (1.5, 2, 2.5) | (1,1,1) | (0.4, 0.5, 0.6) |
| C13 | (2, 2.5, 3) | (0.5, 0.6, 1) | (1, 1.5, 2) | (0.3, 0.4, 0.5) | (0.5, 0.6, 1) | (0.3, 0.4, 0.5) | (1.5, 2, 2.5) | (1,1,1) |

Table 12: Pairwise comparison of Sharing category

| | P3 (Sharing) | | | |
|---|---|---|---|---|
| | C14 | C15 | C16 | C17 |
| C14 | (1,1,1) | (0.4, 0.5, 0.6) | (0.3, 0.4, 0.5) | (1.5, 2, 2.5) |
| C15 | (1.5, 2, 2.5) | (1,1,1) | (0.5, 0.6, 1) | (0.5, 0.6. 1) |
| C16 | (2, 2.5, 3.0) | (1, 1.5, 2) | (1,1,1) | (1, 1.5, 2) |
| C17 | (0.4, 0.5, 0.6) | (1, 1.5, 2) | (0.5, 0.6. 1) | (1,1,1) |

$I_{max} = 4.27$, CI = 0.0856, CR = 0.095

Table 13: Pairwise comparison of Culture category

| | P4 (Culture) | | | | |
|---|---|---|---|---|---|
| | C18 | C19 | C20 | C21 | C22 |
| C18 | (1,1,1) | (0.4, 0.5, 0.6) | (1.5, 2, 2.5) | (0.3, 0.4, 0.5) | (0.4, 0.5, 0.6) |
| C19 | (1.5, 2, 2.5) | (1,1,1) | (2, 2.5, 3) | (0.5, 1, 1.5) | (1, 1.5, 2) |
| C20 | (0.4, 0.5, 0.6) | (0.3, 0.4, 0.5) | (1,1,1) | (2, 2.5, 3) | (2.5, 3, 3.5) |
| C21 | (2, 2.5, 3) | (0.6, 1, 2) | (0.3, 0.4, 0.5) | (1,1,1) | (0.5, 0.6, 1) |
| C22 | (1.5, 2, 2.5) | (0.5, 0.6, 1) | (0.2, 0.3, 0.4) | (1, 1.5, 2) | (1,1,1) |

$I_{max} = 5.85$, CI = 0.21, CR = 0.09

Table 14: Pairwise comparison of in between the categories

| | Automation | Measurement | Sharing | Culture |
|---|---|---|---|---|
| Automation | (1,1,1) | (2, 2.5, 3) | (1, 1.5, 2) | (1.5, 2, 2.5) |
| Measurement | (0.3, 0.4, 0.5) | (1,1,1) | (0.3, 0.4, 0.5) | (0.5, 1, 1.5) |
| Sharing | (0.5, 0.6, 1) | (2, 2.5, 3) | (1,1,1) | (1, 1.5, 2) |
| Culture | (0.4, 0.5, 0.6) | (0.6, 1, 2) | (0.5, 0.6. 1) | (1,1,1) |

$I_{max} = 4.0162$, CI = 0.0053865, CR = 0.0059850

**Step-5: Local and global weight calculation**

The local and global weigh of the challenges and their respective category were determined. The determined results are presented in Table 15, which shows significance of a challenge with in their respective category (local weight) and compared with all the investigated challenges (global weight).

The local weight were determined using the pairwise comparison conducted in step-4. For example, Table 15 shows that the local weight (LW) C3 (Resistance to change, W=0.382099) is the highest challenging factor in the 'Automation' category. It is also observed that C2 (Moving from legacy infrastructure to microservices, W=0.362363) and C5 (No Continuous software engineering center of excellence, W=0.170320) are standout as the second and third most significant challenging factors, respectively.

Moreover, the global weight of each challenge was calculated by multiplying its local weight with the weight of its corresponding category. For example, the global weight (GW) of challenge, C1= 0.049895×0.482232 =0.024061, where 0.482232 is the weight of its category (i.e., automation) and 482232 is its local weight (Table 15). By considering the same process, the global weight (GW) for all the enlisted challenges were determined (Table 15). The presented results (Table 15) show that C3 (Resistance to change, W=0.184261) is ranked as 1st significant challenge for successful execution of Continuous software engineering practices in a software organization. The final ranking of the challenges were determined using the global weights presented in Table 15

Table 15: Determine the global weight of the challenges

| Category | "Category Weight" | Challenges | "Local weight" | "Local rank" | "Global weight (GW)" | "Global rank" |
|---|---|---|---|---|---|---|
| Automation | 0.482232 | C1 | 0.049895 | 4 | 0.024061 | 8 |
| | | C2 | 0.362363 | 2 | 0.174743 | 2 |
| | | C3 | 0.382099 | 1 | 0.184261 | 1 |
| | | C4 | 0.035323 | 5 | 0.017034 | 10 |
| | | C5 | 0.170320 | 3 | 0.082134 | 5 |
| Measurement | 0.135067 | C6 | 0.131098 | 1 | 0.017707 | 9 |
| | | C7 | 0.118479 | 5 | 0.016003 | 14 |
| | | C8 | 0.076133 | 9 | 0.010283 | 18 |
| | | C9 | 0.121058 | 3 | 0.016351 | 12 |
| | | C10 | 0.118755 | 4 | 0.016040 | 13 |
| | | C11 | 0.122095 | 2 | 0.016491 | 11 |
| | | C12 | 0.112572 | 6 | 0.015205 | 15 |
| | | C13 | 0.106366 | 7 | 0.014367 | 16 |
| Sharing | 0.368926 | C14 | 0.17156 | 3 | 0.063293 | 6 |
| | | C15 | 0.23791 | 2 | 0.087771 | 4 |
| | | C16 | 0.43009 | 1 | 0.158672 | 3 |
| | | C17 | 0.16044 | 4 | 0.059191 | 7 |
| Culture | 0.013774 | C18 | 0.092966 | 5 | 0.0012805 | 17 |
| | | C19 | 0.287475 | 1 | 0.0039596 | 19 |
| | | C20 | 0.261335 | 2 | 0.0035996 | 20 |
| | | C21 | 0.186458 | 3 | 0.0025683 | 21 |
| | | C22 | 0.171765 | 4 | 0.0023659 | 22 |

***Step-6:*** *Prioritizing of challenges*

The ultimate objective of fuzzy AHP analysis is to prioritize the investigated challenges concerning to their significance of Continuous software engineering paradigm. The determined final ranking for each challenge is given in Table 16. For determining the final rankings of challenges, the global weights are used. Considering the absolute rankings given in Table 15, C3 (Resistance to change) is the most significant challenge that needs to be addressed for the successful implementation of Continuous software engineering practices in software organizations. Furthermore, it is also observed that C2 (Moving from legacy infrastructure to microservices) and C16 (Communication and Collaboration issues) are declared as the 2nd and 3rd most priority challenges for the implementation of Continuous software engineering practices, respectively. We further noted that C22 ("Heterogeneity in development and operational structure") ranked as least significant challenge for Continuous software engineering paradigm.

Table 16: List of challenges in priority order

| Sr.No. | Challenges | Global rank |
|---|---|---|
| C3 | Lack of Continuous software engineering metrics | 1 |
| C2 | No Continuous software engineering center of excellence | 2 |
| C16 | Communication and Collaboration issues | 3 |
| C15 | Lack of trust relationship | 4 |
| C5 | Building and maintaining the deployment pipeline | 5 |

| C14 | Resistance to change | 6 |
| C17 | Disintermediation of roles within teams | 7 |
| C1 | Moving from legacy infrastructure to microservices | 8 |
| C6 | Overcoming the Dev versus Ops mentality | 9 |
| C4 | Lack of service virtualization | 10 |
| C11 | Configuration Management | 11 |
| C9 | Traceability across the Continuous software engineering landscape | 12 |
| C10 | Artifact management issues | 13 |
| C7 | Lack of integrated tools architecture | 14 |
| C12 | Resources accountability issues | 15 |
| C13 | Incident handling issues | 16 |
| C18 | Inconsistent environments | 17 |
| C8 | Continuous software engineering and regulatory compliance | 18 |
| C19 | Lack of feedback and bugs prioritization | 19 |
| C20 | Lack of flexibility due to rigid Industrial constraints | 20 |
| C21 | Lack of strategic suggestions from the leadership | 21 |
| C22 | Heterogeneity in development and operational structure | 22 |

## 5. Discussion and summary

The ultimate aim of this study is to identify and rank the factors that could negatively affect Continuous software engineering practices. The address the study objectives, the systematic literature review study has been conducted to determine the Continuous software engineering challenging factors reported in the literature and were mapped in the core categories of CAMS model, i.e. ("Culture", "Automation", "Measurement", and Sharing). The challenges and their classification were further verified with expert via questionnaire survey study . Finally, the fuzzy-AHP was performed to prioritize the reported challenges with respect to their significance for the success and progression of Continuous software engineering implementation in software development organizations.

### 5.1 RQ1 (Investigation of Challenges)

The systematic literature review was performed to investigate the Continuous software engineering challenges reported in the literature .

A total of 78 studies were selected by considering the step by step protocols of the SLR approach. The selected studies were explored, and a total of 22 challenges that are critical for the implementation of Continuous software engineering practices were identified. Moreover, the investigated challenges were further classified in the core categories of the CAMS model. The classification of investigated challenges is used for the application of the fuzzy-AHP process.

### 5.2 RQ2 (Investigation of questionnaire survey study)

The questionnaire survey study was conducted to get the insight of the industry experts concerning the findings of the literature review. The results and analysis show that the investigated challenges of Continuous software engineering practices are related toindustry practices.

### 5.3 RQ3 (Prioritization of investigated challenges)

To rank the investigated challenges and their categories, the fuzzy AHP analysis was performed. The pairwise comparison was conducted with the reported of the reported challenge and their categories. To determine the final ranking, the calculated global rank was used. The FAHP technique provides a complete understanding of decision-making problems that consider the Continuous software engineering challenges and their associated categories. Though we calculate the ranks of the identified challenges, and the results are presented in Tables 15 and 16. The results (Table 16) show that C3 (Resistance to change) is the most significant challenge that needs to be addressed for the successful implementation of Continuous software engineering practices in software organizations. Furthermore, it is also observed that C2 (Moving from legacy infrastructure to microservices) and C16 (Communication and Collaboration issues) are declared as the second and third most significant challenges for the implementation of Continuous software engineering practices, respectively.

### 5.4 RQ4 (Prioritization based taxonomy of Continuous software engineering challenges)

The taxonomy of the investigated challenges was developed considering the local and global weights. The challenges were mapped in the core categories of CAMS model, i.e. ("Culture, Automation, Measurement, and Sharing")(Guthrie 2019).

The developed taxonomy (Figure 8), shows that Automation, CW=0.4822) is the top ranked category of reported challenges. This indicated that the experts consider the automationas key area that needs to be focused by the industry experts for the successful execution of Continuous software engineering practices.
Furthermore, it is noted that (Sharing, CW=0.368926), and (Measurement CW=0.135067) are declared as the 2$^{nd}$ and 3red most important categories of the reported Continuous software engineering challenges.

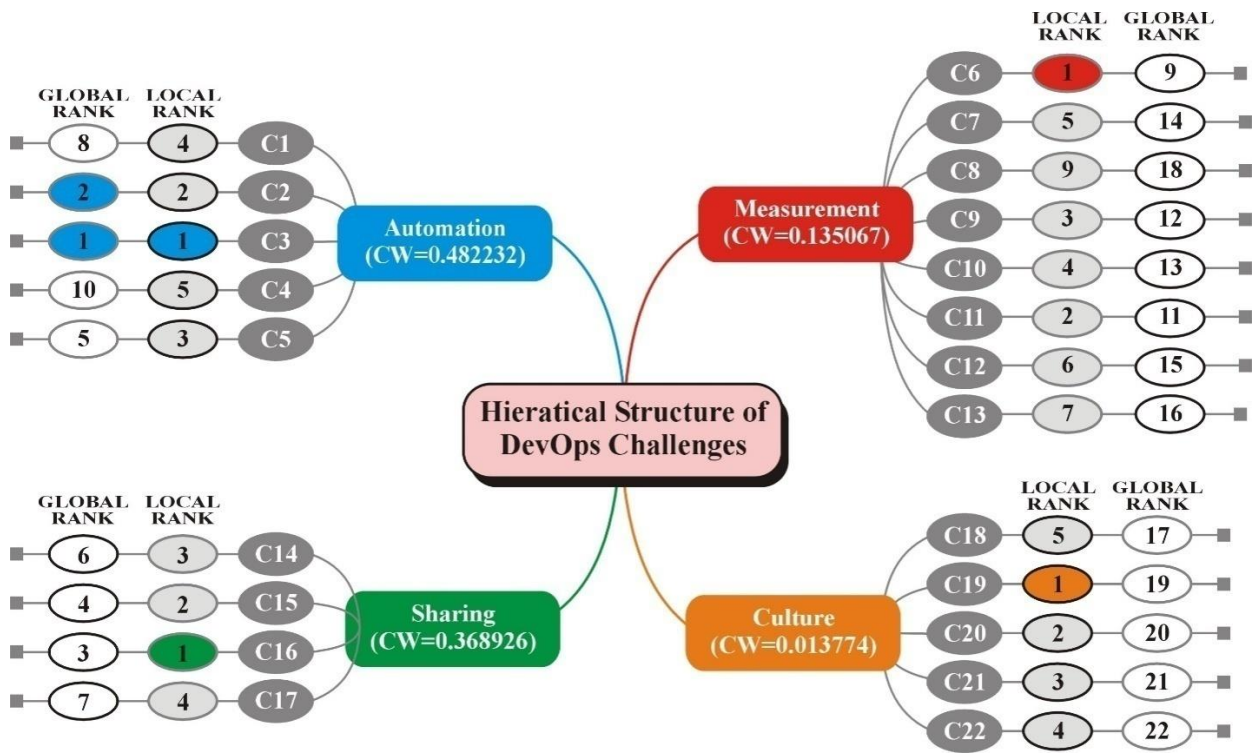

Figure 13: Prioritization based taxonomy of the identified challenges

The developed taxonomy (Figure 13) presents the local and global weights of each challenge. This indicated that how a challenge affects Continuous software engineering activities. We observed that in the automation category,C3 (Resistance to change) is locally ranked as 1st ranked challenge. Consequently, C3 is also ranked as the highest priority challenge for the successful execution of Continuous software engineering practices. Besides, it is observed that C19 (Lack of feedback and bugs prioritization) is ranked as the highest priority challenge in the 'Culture' category, and it's standout the 19th priority challenging factor concerning the global ranking. This renders the importance of C19 within their category and for the overall project.
Similarly, both priority ranks of each challenging factors are presented in the developed taxonomy (Figure 13), which assists the practitioners and researchers to consider the most critical challenges concerning their interest and requirements.

### 5.5 Summary of Research Questions
To address the aim of this study, the factors that could negatively affect the implementation of Continuous software engineering practices in the software development industry are identified and verified with experts via a questionnaire survey study. Further, the fuzzy-AHP technique was applied to determine the rank order of the investigated challenging factors concerning their significance of Continuous software engineering practices. The summary of the research questions findings are presented in Table 17.

Table 17: Summary of findings

| Research Questions | Findings |
|---|---|
| RQ1 | The identified challenges are:<br>Moving from legacy infrastructure to microservices; No Continuous software engineering center of excellence; Lack of Continuous software engineering metrics; Lack of service virtualization; Building and maintaining the deployment pipeline; Overcoming the Dev versus Ops mentality; Lack of integrated tools architecture; Continuous software engineering and regulatory compliance; Traceability across the Continuous software engineering landscape; Artifact management issues; Configuration Management; Resources accountability issues; Incident handling issues; Resistance to change, Lack of trust relationship; Communication and Collaboration issues; Disintermediation of roles within teams; Inconsistent environments; Lack of feedback and bugs prioritization;"Lack of flexibility due to rigid Industrial constraints"; Lack of strategic suggestions from leadership;" Heterogeneity in development and operational structure". |
| RQ2 | The majority of the survey respondents are agreed as the identified challenges from the literature review study, and their classification is related to the industrial practices. |
| RQ3 | Using the fuzzy-AHP method, the explore list of Continuous software engineering challenges and their categories are prioritized. According to the results Automation, CW=0.4822) is the highest-ranked category of the investigated challenges and C3 (Resistance to change), C2 (Moving from legacy infrastructure to microservices) and C16 (Communication and Collaboration issues) are declared most significant challenges for the Continuous software engineering practices in software organizations. |
| RQ | Both priority ranks (local and global) of each challenge are presented in Figure 13, which assists the industry experts and academic researchers to consider the most critical problem concerning their interest and for the successful implementation of Continuous software engineering practice in software organizations. |

## 6. Threats to Validity

Some potential risks need to be fixed for the generalization of study results. For example, there may be the researcher's biasness in the literature findings. We have conducted an inter-rater reliability test to check the researcher's biasness, and the results show that the findings are consistent and unbiased.

An external threat towards the generalization of study results is the small sample size of empirical study. The data set consists (n=93) might not be strong enough to generalize the results of this study. Though, by considering the existence of other software engineering domains, this sample size is representative of generalizing the study results (Niazi, Mahmood et al. 2015, Khan, Keung et al. 2017, Khan and Akbar 2019).

Most of the survey respondents were from developing countries (Asian countries); this may hinder to generalize the study results. Moreover, we also noted that a representative number of respondents are form developed continents (the USA or Australia), and this allows the generalization of results.

## 7. Study Implication

The study sheds light on the challenging factors of Continuous software engineering implementation in the software industry, reported by the researchers and practitioners. The detailed overview of the Continuous software engineering existing literature and empirical investigations will provide the body of knowledge to researchers and practitioners to develop effective plans and strategies for the success and progression of the Continuous software engineering paradigm.

Moreover, the fuzzy AHP approach was performed to rank the reported challenges and their categories considering their significance for the successful implementation of Continuous software engineering activities. The calculated ranks orders serve as a knowledge base for practitioners and researchers to consider the most critical challenging factors or priority basis.

Besides, this study provides a taxonomy of the challenging factor by considering their global and local priorities. The identified challenging factors were classified into four key categories, and each category presents a particular key knowledge area of Continuous software engineering process improvement. The challenges of each category contain local and global weights that assist researchers and practitioners in choosing the most significant challenging factor concerning their interest and working area".

## 8. Conclusion and Future directions

It is the priority of every business organization to get a good return on investment; therefore, the software development industry continuously looking the ways to develop effective development approaches. The Continuous software engineering is the latest and most significant approach, and it provides more satisfactory results. The significance of the Continuous software engineering process, motivate us to explore the challenging factors faced by the practitioners while adopting the Continuous software engineering process.

The systematic literature review approach has been adopted to identify the challenges of Continuous software engineering practices. The identified challenges were further mapped into core categories and verified with experts using a questionnaire survey approach. The empirical results show that the identified Continuous software engineering challenges are related to industry practices. This renders that it is critical to address the identified challenges for the successful implementation of Continuous software engineering practices.

Moreover, the fuzzy-AHP technique was applied to prioritize the investigated challenges and their categories with respect to their significance for the implementation of Continuous software engineering practices. The local and global ranks were determined using the fuzzy-AHP approach. The local ranks present the priority order of a challenge within their particular category. The global ranks show the significance of Continuous software engineering challenges for the overall study objective. By considering the final rankings, Automation is declared as the highest ranked category of Continuous software engineering challenges. . C3 (Resistance to change), C2 (Moving from legacy infrastructure to microservices), and C16 (Communication and Collaboration issues) are declared the most significant challenges for the Continuous software engineering paradigm in software industry. Study findings also provide the prioritization based taxonomy of the investigated challenges, which assists the researchers and practitioners in developing the effective strategies for the success and progression of Continuous software engineering practices.

In the future, we will conduct the multivocal literature study to investigate the factors that have a negative and positive impact of Continuous software engineering practices. We also plan to perform empirical research to identify success factors and challenges. Besides, we also conducted a literature review and an empirical study to explore the best practices for the success and progression of Continuous software engineering practices.

**The appendixes are given in the following links:**
Appendix-A: "List of selected studies and their quality assessment score (https://tinyurl.com/tjw89aj)"
Appendix-B: "Questionnaire survey sample (https://tinyurl.com/quo3etw)"
Appendix-C: "Sample of pairwise comparison questionnaire (https://tinyurl.com/u7qxo7x)"

**"Compliance with ethical standards:"**
"Conflict of interest: All authors declare that there is no conflict of interest".